\documentclass[aps,twocolumn]{revtex4}
\usepackage{epsf}
 
\begin{document}

\title{Energy harvesting in the nonlinear piezoelastic systems driven by stochastic excitations}

\author{K. Kucab, G. G\'orski, and J. Mizia}

\affiliation{ Faculty of Mathematics and Natural Sciences, University of Rzesz\'ow, 35-310 Rzesz\'ow, Poland}

\begin{abstract}
In our work we examine the influence of background stochastic excitations on the power output generated by the energy harvester. The harvesting device we considered is composed of two small magnets, attached to piezoelastic oscillators, which are moving in the magnetic field produced by the static magnets fastened directly to the device. The moving magnets are coupled together by an electrical circuit.  The nonlinearity of the system is achieved by the variable distance between the moving and static magnets, and additionally by the mass difference between them. The stochastic noise applied to the system obeys the Gaussian distribution. We examine the power output generated on the electric load at different excitation frequencies for different values of mistuning parameters and different stochastic amplitudes. We also present the mean value of the power generated averaged over a range of considered frequencies. 

\end{abstract}

\maketitle

In recent years, the energy harvesting devices based on electromechanical systems have been widely studied. Due to the fact that the electronic devices are characterized by very small power consumption, these systems seems to be very important energy sources, especially for devices using the piezoelectric effect. We take into account that energy harvesting devices should have a broad band of frequencies for which the power generated is relatively large compared to the ‘simple’ case, when the value of power obtained is high only at one resonance frequency. This problem of obtaining a broad spectrum of frequencies with high power output is very important for energy harvesting systems.

There have been many attempts to obtain energy harvesting devices with an extended frequency bandwidth, and many theoretical and experimental papers devoted to this topic. The set of parallel single degree of freedom harvesters tuned at slightly different resonant frequencies \cite{Ref1}, or the piezoelectric multifrequency energy converter for power harvesting in autonomous microsystems \cite{Ref2}, were studied. There were attempts to combine several beams capable of producing energy from the vibrations in one device. There were also attempts to consider a harvester consisting of a serial set of two beams connected to each other to form an L-shape \cite{Ref3} and investigations into the relationship between two piezoelectric harvesters \cite{Ref4}. Actually, the dynamic systems with attached magnets working in the nonlinear regime have shown the possibility of generating broadband harvesting solutions \cite{Ref5}-\cite{Ref11}. There are also attempts to improve the models of energy harvesting devices by introducing the stochastic behavior of the environment. Vocca et al. \cite{Ref12} introduced the stochastic term in the equation of motion for a moving magnet. The noise they considered was obtained in the experimental way (the data was collected from the analysis of train, car and microwave oven). In turn, Martens et al. \cite{Ref13} studied the efficiency of vibration energy harvesting systems with stochastic ambient excitations by solving corresponding Fokker-Planck equations.

In our paper we take into account the recent findings about the role of nonlinearity in energy harvesting problems. We introduce two parameters to obtain nonlinearity, i.e. the variable distance between moving and stable magnets, and different vibrating masses described by the mass disproportion parameter. Additionally, we include the stochastic term, which allows us to consider the system in a more realistic way.

\section{The Model}
\label{sec:2}

\begin{figure}
\epsfxsize=6cm{\epsffile{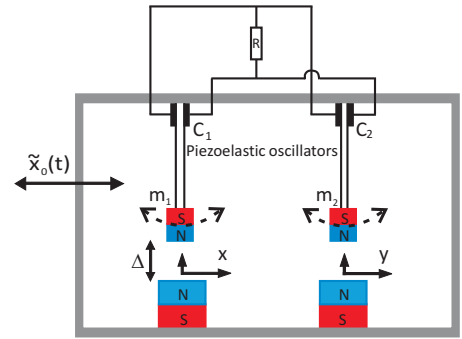}}
\caption{Schematic of the device considered in the model. Full description in the text.}
\label{fig:1}       
\end{figure}

In our work we study the system in which two small magnets attached to piezoelastic oscillators are moving in the magnetic field produced by the static magnets attached to the device. The moving magnets are coupled to each other by an electrical circuit. The whole device is subjected to vibrations, which are assumed may have a harmonic and stochastic origin. The schematic diagram of the model is presented in Figure \ref{fig:1}.

The moving magnets, characterized by their masses $m_1$ and $m_2$, are attached to the piezoelastic beams and are separated from the static magnets by an air gap of thickness $\Delta$. The value of distance $\Delta$ is responsible for creating the bistable state of their potential energy. The masses vibrate in the quasi-horizontal direction described by the variables $x$ and $y$, which indicate their displacements from their initial positions. The distance between static magnets should be large enough to avoid mutual influence of the moving magnets with each other and also avoid the interaction between the moving magnet and the static field acting on the neighboring moving magnet. The mass disproportion parameter $\alpha$ is defined by the relation: $\alpha=m_2/m_1$.

\noindent
The stochastic kinetic excitation $\tilde x_0 \left( t \right)$ is applied to the harvester (we use the notation, where the tilde sign over the symbol indicates the stochastic behavior of a corresponding physical quantity). The energy generated by the piezoelastic beams is received by the resistance $R$ coupled with the device by an electrical circuit containing two capacitances $C_1$ and $C_2$. The electrical circuit can be characterized by the time constant $\tau  = R\left( {C_1  + C_2 } \right)$.

The physics of the system is described in terms of the differential equations which couple the mass displacements $x$, $y$ of moving magnets, and the voltage $V$ induced on the electric load. The dynamics of the tip mass motion is described by the Newton's law 

\begin{equation}
\left\{ \begin{array}{l}
 \,\,\,\,m_1 \ddot x\left( t \right) =  - \frac{{dU\left( {x,\Delta } \right)}}{{dx}} - \gamma \dot x - \chi V\left( t \right) + \tilde F\left( t \right) \\ 
 \alpha m_1 \ddot y\left( t \right) =  - \frac{{dU\left( {y,\Delta } \right)}}{{dy}} - \gamma \dot y - \chi V\left( t \right) + \tilde F\left( t \right) \\ 
 \end{array} \right.
\label{Eq:1}
\end{equation}
with the equation for the electrical potential $V(t)$  
\begin{equation}
\dot V\left( t \right) =  - \frac{1}{\tau }V\left( t \right) + \kappa \dot x + \kappa \dot y,
\label{Eq:2}
\end{equation}

\noindent
where the over-dot sign means the time derivative, $\gamma$ is the viscous parameter, $\chi$ is the piezoelectric coupling constant, and $\kappa$ is the position to voltage coupling coefficient. The inertial force, $\tilde F\left( t \right)$, has the following form

\begin{equation}
\tilde F\left( t \right) = \tilde F_0 \left( t \right)\sin \omega t,
\label{Eq:3}
\end{equation}

\noindent
where the total amplitude $\tilde F_0\left( t \right)$ is given by the relation

\begin{equation}
\tilde F_0 \left( t \right) = F_0  \left[ {1 + \tilde \delta \left( t \right)} \right].
\label{Eq:4}
\end{equation}

\noindent
The stochastic, time-dependent function appearing in Eq. (\ref{Eq:4}) is equal to $\tilde \delta \left( t \right) = \delta _0  \tilde \varepsilon \left( t \right)$, with $\tilde \varepsilon \left( t \right)$ being the set of randomly generated numbers (with mean value equal to 0, and the standard deviation equal to 1) obeying the Gaussian distribution. The normalization factor $\delta_0$ transforms the system from the non-stochastic regime ($\delta_0=0$) to the mixed one ($\delta_0\ne 0$), i.e. when the stochastic excitations are ``added'' to the harmonic ones.
The stochastic force presented above can be realized e.g. by considering the periodically moving parts exposed to the random shakes. For example this can be a situation of the rotating car wheel moving on the uneven road, or the helicopter rotor blades rotating in randomly changing engine conditions.
The angular excitation frequency $\omega$, and amplitude of excitation $F_0$ are treated for simplicity as the parameters. We use the standard relation connecting $\omega$ with the applied frequency $\nu$: $\omega=2\pi \nu$.
The potential energy of moving masses appearing in Eqs (\ref{Eq:1}) is given (after \cite{Ref12}) by

\begin{equation}
U\left( {x,\Delta } \right) =  \frac{1}{2}k_\mathrm{eff} x^2  + \left( {ax^2  + b\Delta ^2 } \right)^{ - 3/2}. 
\label{Eq:5}
\end{equation}

\noindent
The parameters $k_\mathrm{eff}$, $a$ and $b$ represent the physical properties of piezoelastic beams, and the $\Delta$ parameter (see Figure \ref{fig:1}) is responsible for transferring the system from linear regime ($\Delta>15\textrm{mm}$) to the bistable regime \cite{Ref12}.

\section{Numerical results}
\label{sec:3}

We perform the computations with parameters used in \cite{Ref11}, where the corresponding non-stochastic system has been investigated. The values of parameters used in our model are presented in the Table \ref{tab:1}.

\begin{figure}[b]
\epsfxsize=6cm{\epsffile{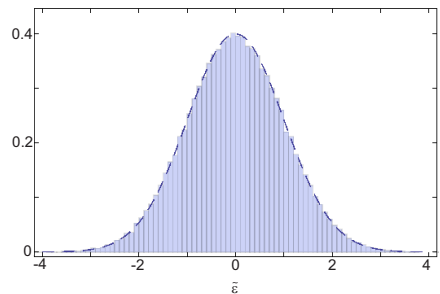}}
\caption{The probability density histogram of 100\,000 random numbers used in our computations. The dashed line presents the normal distribution function with mean value equal to 0 and the standard deviation equal to 1. Full description in the text.}
\label{fig:2}       
\end{figure}

\begin{table}
\caption{The values of parameters used in our model.}
\label{tab:1}       
\begin{tabular}{lll}
\hline\noalign{\smallskip}
Variable & Variable value & Variable unit  \\
\noalign{\smallskip}\hline\noalign{\smallskip}
$\alpha$ & 1.0$\div$1.2 & [1] \\
$\delta_0$ & 0.0$\div$1.0 & [1] \\
$\Delta$ & 5;\,15 & [mm] \\
$\gamma$ & $22 \cdot 10^{ - 3}$ & [kg/s] \\
$\chi$ & $1.1 \cdot 10^{ - 3}$ & [N/V] \\
$\tau$ & $33.6 \cdot 10^{ - 3}$ & [s] \\
$R$ & $300$ & [k$\mathrm{\Omega}$] \\
$\kappa$ & $4.15 \cdot 10^{ 3}$ & [V/m] \\
$m_1$ & $18 \cdot 10^{ -3}$ & [kg] \\
$k_\mathrm{eff}$ & 26.6 & [N/m] \\
$a$ & $2.82 \cdot 10^{ 7}$ & [J$^{-2/3}$m$^{-1}$] \\
$b$ & $3.19 \cdot 10^{ 6}$ & [J$^{-2/3}$m$^{-1}$] \\
$F_0$ & $176.6 \cdot 10^{ -3}$ & [N] \\
\noalign{\smallskip}\hline
\end{tabular}
\end{table}

We will investigate the influence of stochastic excitations on the output power generated for different values of mistuning parameter $\alpha$. As was shown in our previous paper \cite{Ref11}, the decrease of the air gap between moving and stable magnets to values less than $\Delta<5\textrm{mm}$ raises a strong potential barrier which prevents the passage of the magnets through the $x,y=0$ points, i.e. to another side of the potential well, and therefore the value of maximum power generated by the device is much smaller compared to the situation, when the magnet can pass the potential barrier. Therefore, in this work we use the values of parameter $\Delta=5\textrm{mm}$ (bistable case) and $\Delta=15\textrm{mm}$ (monostable case).

We use the set of 100\,000 random real numbers obeying the normal distribution function. The set of random numbers, with mean value equal 0 and the standard deviation equal to 1, presented as the probability density histogram is shown in Figure \ref{fig:2}.

\begin{figure}
\epsfxsize=5cm{\epsffile{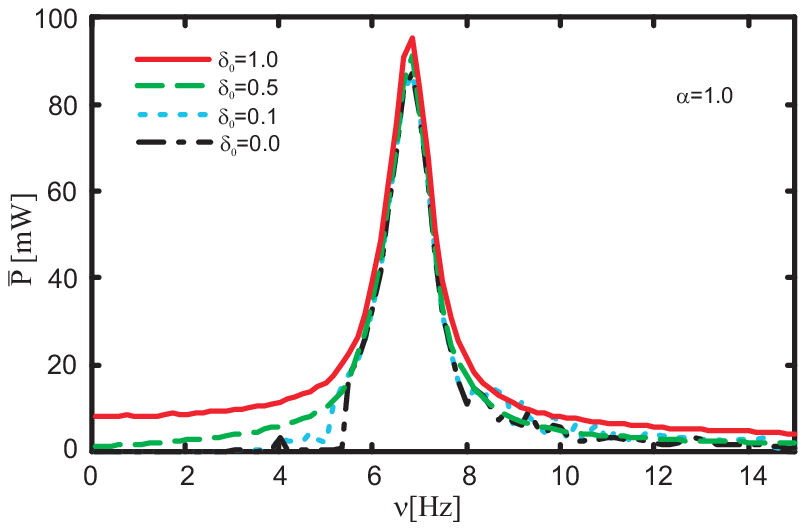}}
\epsfxsize=5cm{\epsffile{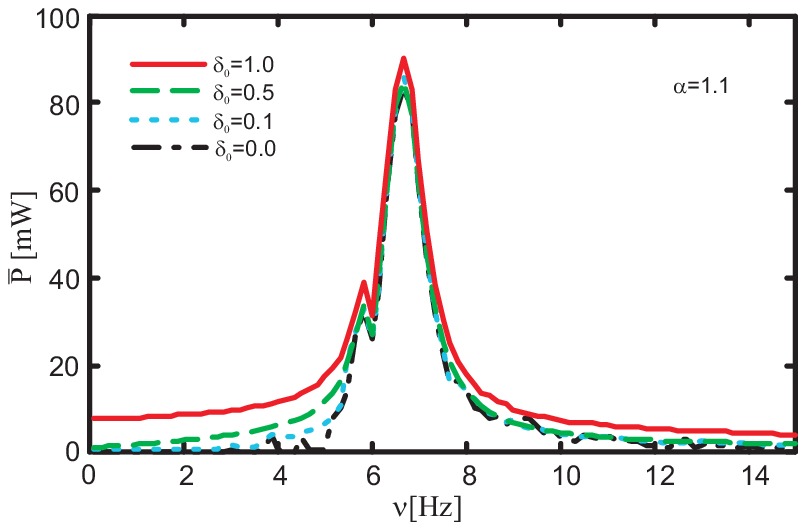}}
\epsfxsize=5cm{\epsffile{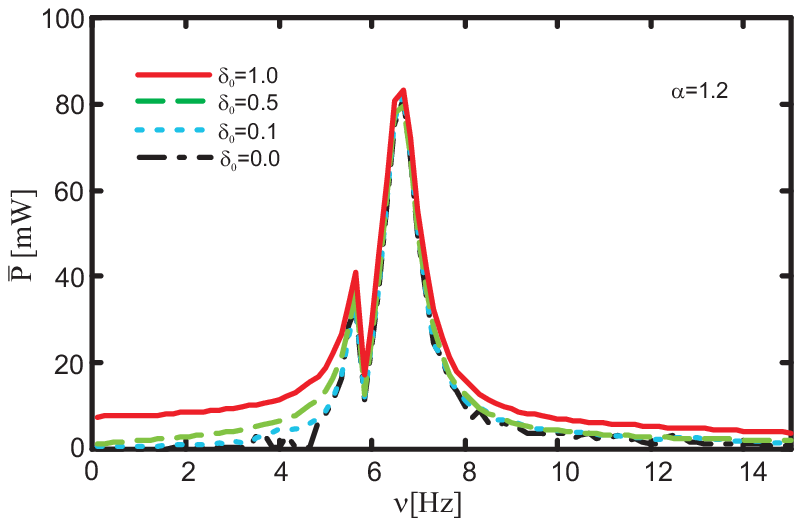}}
\caption{The mean power generated on the electric load $R$ as a function of frequency $\nu$ for different values of mass disproportion parameter $\alpha$, and for different values of $\delta_0$ factor ($\Delta=5\textrm{mm}$).}
\label{fig:3}       
\end{figure}

In Figure \ref{fig:3} we present the mean power as a function of frequency $\nu$ for different values of mass disproportion parameter $\alpha$ and for different values of the $\delta_0$ factor. The mean power is given by the relation $ \bar{P} = \left\langle {V^2 } \right\rangle /R$, where the $\left\langle {V^2 } \right\rangle$ term is the mean squared voltage generated on the electric load $R$.

As we can see in Figure \ref{fig:3}, increasing the $\delta_0$ factor, i.e. introducing the stochastic behavior of the external force, causes an increase of output power, especially in the low frequency regime. 

\begin{figure}
\epsfxsize=6cm{\epsffile{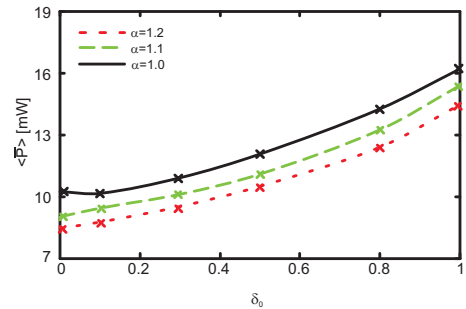}}
\caption{The effective output power generated on the electric load $R$ as a function of parameter $\delta_0$ for different values of mass disproportion parameter $\alpha$ ($\Delta=5\textrm{mm}$).}
\label{fig:4}       
\end{figure}

To compare the efficiency in the broad frequency range for the examples considered, we show in Figure~\ref{fig:4} the effective output power as a function of parameter $\delta_0$ for different values of $\alpha$. The effective output power is obtained as the average of the mean power over the considered excitation frequencies, $\nu$, changing from 0 to 15Hz.

\begin{figure}
\epsfxsize=5cm{\epsffile{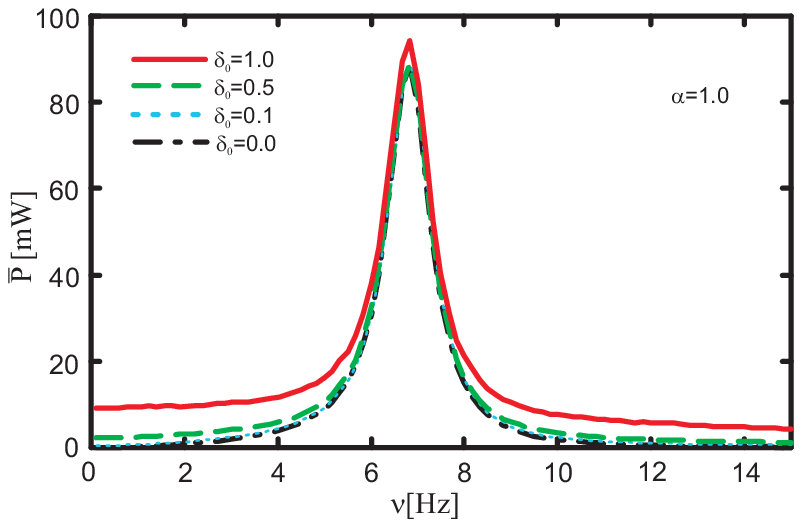}}
\epsfxsize=5cm{\epsffile{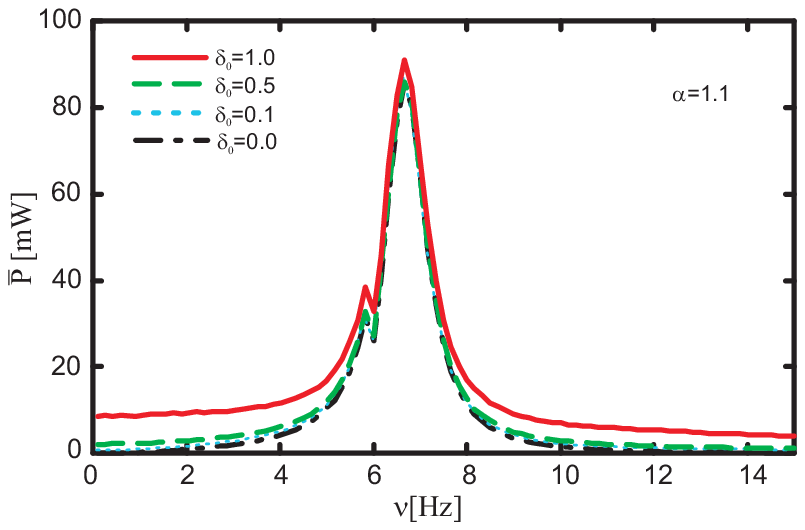}}
\epsfxsize=5cm{\epsffile{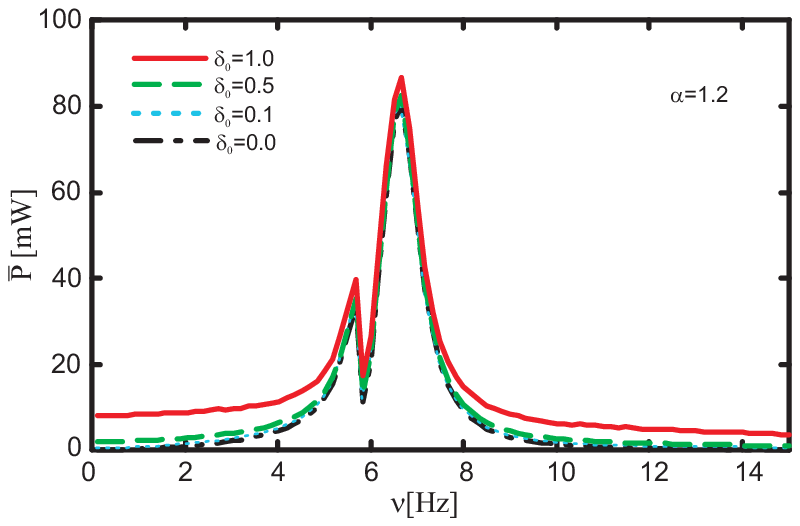}}
\caption{The mean power generated on the electric load $R$ as a function of excitation frequency $\nu$ for different values of mass disproportion parameter $\alpha$, and different values of $\delta_0$ factor ($\Delta=15\textrm{mm}$).}
\label{fig:5}       
\end{figure}

As we can see in Figure~\ref{fig:4} the effective output power grows with increasing parameter $\delta_0$ for all values of mistuning parameter $\alpha$ considered in this paper. The increase of parameter $\alpha$ causes the system to have two resonance frequencies, which are visible in Figure \ref{fig:3}. It also slightly broadens the range of detected power (see Figure 3). On the other hand the total power output decreases with growing mistuning parameter (see Figure \ref{fig:4}). 
For comparison we present in Figure \ref{fig:5} the corresponding results for the parameter $\Delta=15\textrm{mm}$, i.e. when the device works in the linear regime (monostable case). In this regime each  moving magnet, shown in Figure \ref{fig:1}, has only one stable minimum.

\begin{figure}
\epsfxsize=6cm{\epsffile{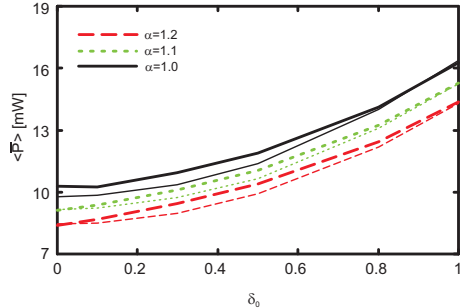}}
\caption{Comparison of the effective output power generated on the electric load $R$ as a function of parameter $\delta_0$ different values of mass disproportion parameter  $\alpha$; thick lines - the bistable case ($\Delta=5\textrm{mm}$); thin lines - the monostable case ($\Delta=15\textrm{mm}$).}
\label{fig:6}       
\end{figure}

The comparison of effective output power as a function of parameter $\delta_0$ for different values of $\alpha$ for the monostable ($\Delta=15\textrm{mm}$) and bistable ($\Delta=5\textrm{mm}$) potential is presented in Figure \ref{fig:6}. As we can see, the results obtained at $\Delta=15\textrm{mm}$ are qualitatively the same as at $\Delta=5\textrm{mm}$. Introducing double minima potential improves the efficiency of the device for almost all considered values of stochastic amplitude $\delta_0$ but only to a small degree.

The computations in this work were done with the Mathematica package. We used the NDSolve numerical differential equation solver. The time simulation to reach the self consistent solution was of the order of one hour when the time interval for sampling the solution was divided to approximately ten million steps. All the average values in numerical integrations were obtained using the Monte Carlo method with the number of steps in the order of one million. The random numbers were generated by the random number generator included in the Mathematica package (the same generator was also used in computing the Monte Carlo integrals).

\section{Conclusions}
\label{sec:4}

We have examined the energy generated by the energy harvesting device attached to a vibrating source in which harmonic excitations are mixed with the stochastic excitations. The results we obtained are compared with a purely harmonic case. The bistability of the system was achieved by the relatively small distance between moving and stable magnets ($\Delta=5\textrm{mm}$), so the harvester worked mostly in the unsynchronized regime. We have analyzed the total mean power generated on the resistance $R$ in the function of excitation frequency in the harmonic and mixed stochastic regimes. The set of initial conditions corresponding to the stable initial positions of the moving magnets at $t=0$ (i.e. the positions for which the potential energy reaches its minimum) was used. 

\noindent
At increasing parameter $\delta_0$ we have observed growth of the generated mean power, especially in the low frequency regime. It is interesting that the maximum power near the resonance frequency remained unchanged. The effective output power was increasing with increasing strength of stochastic excitation at all examined values of mistuning parameter $\alpha$ which by itself did not increase the efficiency of harvester. It was shown that the stochastic behavior of the environment improves the efficiency of the energy harvesting device as compared to the harmonic case.

\begin{acknowledgements}
This work was done due to partial support from Centre for Innovation and Transfer of Natural Sciences and Engineering Knowledge at the Rzesz\'ow University.
We would like to express our gratitude to Prof. Grzegorz Litak for helpful suggestions and discussions.
\end{acknowledgements}



\end{document}